\documentclass[11pt]{article}
\usepackage{epsfig}
\newcommand{\ct}[1]{$^{\cite{#1}}$}

\newcommand{\bee}{\begin{equation}}
\newcommand{\ee}{\end{equation}}
\newcommand{\beea}{\begin{eqnarray}}
\newcommand{\eea}{\end{eqnarray}}

\begin{document}
\title{A Quasi-staggered Scheme on Lattice }
\author{{Da Qing Liu}\\
       {\small Institute of High Energy Physics, Chinese Academy
                of Sciences, P.R. China}\\
       }
\maketitle
\begin{center}
\begin{minipage}{5.5in}
\vskip 0.4in {\bf Abstract} Utilizing a picture of string and
string spinors, we show a simpler version of staggered action. The
advantage of this action is that in this action there always exist
pair of quarks with different masses.
\end{minipage}
\end{center}
\vskip 0.5in \indent

\newpage

\section{Introduction}
There are different actions on the fermion lattice simulation,
such as Wilson action\ct{wilson}, overlapped action\ct{overlap},
SLAC action\ct{slac}. The staggered action, proposed by Kogut and
Susskind\ct{kgt}, stands in these actions. It
is\ct{kgt,stagger,np1}, of course, a very interesting action for
its elegant treatment in solving the fermion doublers. There are
arguments that it is really QCD\ct{rooting}. However, a drawback
of this action is that there are four degenerate fermions in this
action. How to reduce the number of the degenerate fermion in
dimension 4 lattice is a special topic on this action. Besides
this, the spin-isospin symmetry is also an unexpected symmetry of
staggered action, for this symmetry does not occur in standard
continuum QCD.

the modern simulation to decrease the number of fermion in
staggered action is based on the "rooting trick"\ct{rooting}.
However, the validity of "rooting trick" is under discussion. For
instance, a fractional power of the determinant is, in general,
not a legitimate operation in quantum field theory. Furthermore,
the locality  is not obvious in this trick.

Gamma matrix, particularly, $\gamma_i$, can be written as direct
product of two two-by-two matrices, $\tau_A\otimes\tau_i$, where
$\tau_i$ is Pauli matrix and $\tau_A$ is a fixed Pauli matrix,
such as $\tau_1$, $\tau_2$ and $\tau_3$. After $\tau_A$ has been
adopt,  we can set $\gamma_0$ and $\gamma_5$ as $\tau_B\otimes
\mathbf{1}_{2\times 2}$ and $\tau_C\otimes \mathbf{1}_{2\times 2}$
respectively, provided $A\neq B\neq C$. We shall use this
interesting property of gamma matrices and a picture of string
spinors, the notation of which is shown in the context, to make an
attempt to reduce the fermion number in staggered action. The
scheme adopted here is called as quasi-staggered scheme.

Section 2 is a list of results of structure of algebra which will
been used in the scheme. The detail discussions of the
quasi-staggered scheme are shown in section 3. Section 4 shows
some basic properties of string spinor. Then we give a summary in
section 5.

\section{The algebraic structure of the two-component theory}
Consider a four dimensional lattice theory with lattice spacing
$a_s=a_t=1$. We assume that the lattice, which is divided into
$N^4$ grids, has a periodic conditions here for simplification.

On this lattice we have $N^4$ string $n=(n_0,n_1,n_2,n_3)$, in
which we often define four points $x(n)=(x_0,x_1)$. For these four
points $x$'s, $(0,0)$, $(1,0)$, $(0,1)$ and $(1,1)$ stand for
points \(n\), \(n+b_0\), \(n+b_1\) and \(n+b_0+b_1\) respectively,
where $b_0=h{a_0\over 2}$ and $b_1=\sum\limits_{i=1}^3 h{a_i\over
2}$ (At the moment we set parameter $h=1$). To distinguish points
belonging to the same string sometimes we also use $(n,x)$ to
represent this points, for instance, $(n,(1,0))$ stands for
$n+b_0$. Two-component spinors/tastes $\varphi(x)$ and
$\bar{\varphi}(x)$ are defined on each of the four points.


Furthermore, \bee c_0(x)=1,\,c_1(x)=(-1)^{x_0} \label{c01}\ee are
also defined on the link between points $x$ and $x+\vec{x}_0$ or
between points $x$ and $x+\vec{x}_1$ (There are ambiguities in
$c_1$ when $x_1=1$. In fact, all the variable links between points
$n+b_1$, i.e. (n,(0,1)), and $n+a_1$, $n+a_2$ or $n+a_3$ are
described as $c_1((0,1))$. The case is  similar for $c((1,1))$.
The importance is that $c_1$ does not depend on $x_1$. We shall
show that this ambiguity is irrelevant in the next section).

We also define a set of linear operators $\Gamma_\rho$ ($\rho=1$
or 2) which transform the subspace of functions $\varphi(x)$
associated to sting $n$ into itself: \beea
\Gamma_\rho\varphi(x)=c_\rho(x)\varphi(x+x_\rho), &&
\mbox{$x_\rho$ even}, \nonumber \\
\Gamma_\rho\varphi(x)=c_\rho(x-x_\rho)\varphi(x-x_\rho), &&
\mbox{$x_\rho$ odd}, \eea or \bee
\Gamma_\rho\varphi(x)=c_\rho(x)\varphi(x+(-)^{x_\rho}x_\rho). \ee

Immediately, we have \bee \Gamma_\rho^2={\bf 1},\,\,
\{\Gamma_\rho,\Gamma_\sigma\}=2\delta_{\rho\sigma}. \ee

We also define 2-by-2 matrices, for instance,
\bee \lambda_1=\tau_1=\left(%
\begin{array}{cc}
  0 & 1 \\
  1 & 0 \\
\end{array}%
\right),\,\lambda_0=\tau_3=\left(%
\begin{array}{cc}
  1 & 0 \\
  0 & -1 \\
\end{array}%
\right),\,\lambda_5=\tau_2=\left(%
\begin{array}{cc}
  0 & -i \\
  i & 0 \\
\end{array}%
\right). \ee This definitions are just for convention, other
sequence of Pauli matrices is also suitable. Correspondingly,
gamma matrices are defined as \bee
\gamma_i=\tau_1\otimes\tau_i=\left(%
\begin{array}{cc}
  0 & \tau_i \\
  \tau_i & 0 \\
\end{array}%
\right),\, \gamma_0=\tau_3\otimes {\bf 1}=\left(%
\begin{array}{cc}
  {\bf 1} & 0 \\
  0 & {\bf -1} \\
\end{array}%
\right),\,\gamma_5=\tau_2\otimes {\bf 1}=\left(%
\begin{array}{cc}
  0 & -i{\bf 1} \\
  i{\bf 1} & 0 \\
\end{array}%
\right). \ee

Choose an arbitrary path $b_{\mu_1}+b_{\mu_2}+\cdots+b_{\mu_k}$
joining $x^I(n)=(0,0)$ to $x(n)$, i.e. \bee
x=x_{\mu_1}+x_{\mu_2}+\cdots+x_{\mu_k}, \ee we define $p(x)$ as
\bee p(x)=c_{\mu_1}(x^I)\lambda_{\mu_1}
c_{\mu_2}(x^I+x_{\mu_1})\lambda_{\mu_2}+\cdots
+c_{\mu_k}(x-x_{\mu_k})\lambda_{\mu_k}. \ee Explicitly, \bee
p((0,0))={\bf 1},\,p((1,0))=\lambda_0,\,p((0,1))=\lambda_1, \,
p((1,1))=-\lambda_0\lambda_1=-i\lambda_5. \ee It is easy to verify
that \bee p(x)^{\dag \alpha}_\beta
p(y)^\beta_\alpha=2\delta_{xy},\, \lambda_\rho \lambda_5 p(x)
\lambda_5 \lambda_\rho=(-)^{x_\rho}p(x). \ee

The algebra listed here is in fact a 2-dimensional version of the
4-dimensional algebra shown in reference \cite{stagger}. Therefore
we only list the results here, For the detail discussions,
especially for the discussions in the case of four dimension, we
refer to reference \cite{stagger}.

\section{Formulae of the staggered action}

Before the discussions of staggered action we should code the four
 $\varphi(x)$'s into a matrix $\Psi$ associated to
\(n\), \bee \Psi^\alpha_\beta(n)={1\over
\sqrt{2}}\sum\limits_{x\in n} \varphi(x)p^{\dag
\alpha}_\beta(x),\, \bar{\Psi}^\alpha_\beta(n)={1\over
\sqrt{2}}\sum\limits_{x\in n} p^\alpha_\beta(x)\bar{\varphi}(x).
\ee $\Psi$'s are called string spinor thereinafter. Then \bee
\varphi(x)={1\over \sqrt{2}}\Psi^\alpha_\beta(n)p^\beta_\alpha(x)
,\, \bar{\varphi}(x)={1\over
\sqrt{2}}\bar{\Psi}^\alpha_\beta(n)p^{\dag \beta}_\alpha(x). \ee

It should be emphasized here that $\Psi(n)$ is a $4\times 2$
matrix since $\varphi(x)$'s are two-component spinors. The
explicit forms of $\Psi$ and $\varphi(x)$ are  \bee
\Psi(n)={1\over \sqrt{2}}\left(%
\begin{array}{cc}
  \varphi((0,0))+\varphi((1,0)) & \varphi((0,1))+\varphi((1,1)) \\
  \varphi((0,1))-\varphi((1,1)) & \varphi((0,0))-\varphi((1,0)) \\
\end{array}%
\right), \ee and \bee \varphi((0,0))={\Psi^1_1+\Psi^2_2\over
\sqrt{2}},\, \varphi((1,0))={\Psi^1_1-\Psi^2_2\over \sqrt{2}},\,
\varphi((0,1))={\Psi^1_2+\Psi^2_1\over \sqrt{2}},\,
\varphi((1,1))={\Psi^1_2-\Psi^2_1\over \sqrt{2}} \ee respectively.
For $\bar{\Psi}$ and $\bar{\varphi}$ the formulae are similar.

To translate the two-component theory into ordinary algebra of
four-component spinors $\psi_i,\, i=1,2$, we define
$\psi_i=\Psi^{\bf \cdot}_i$ and $\bar{\psi}^i=\bar{\Psi}^i_{\bf
\cdot}$. Therefore, \beea
\bar{\psi}^i(n)\psi_i(n)&=&\bar{\Psi}^i_j\Psi^j_i=
\sum\limits_{x\in n}\bar{\varphi}(x)\varphi(x), \nonumber \\
\bar{\psi}^i(n)\psi_j(n)\lambda_{0ji}&=&
\bar{\Psi}^i_k\Psi^k_j\lambda_{0ji}=\sum\limits_{x\in n}
\bar{\varphi}(x)\varphi(x+(-)^{x_0}\vec{x}_0)(-)^{x_1}.
\label{mss}
 \eea

This equation is crucial. As we know, staggered action introduces
spinor$\otimes$taste interactions, which will lead to a
spontaneous breaking of taste symmetry. However, these
interactions are at higher order and we expect the symmetry
breaking is not large, i.e. all the considered tastes are almost
degenerate. In other words, it is difficult in standard staggered
action to simulate quarks with different masses. If we ignore the
spinor$\otimes$taste interactions and weak interactions which may
lead to admixture of different quarks, mass matrix $M$ with two
taste will have a form of diagonal 2-by-2 matrix. Suppose $M$ can
be decomposed as $M=m_0{\bf 1}_{2\times 2}+m_1\lambda_0$,
 equations in (\ref{mss}) supplies a obvious way to simulate quarks
with different masses, for the second equation in (\ref{mss}) can
produce mass splitting of the two quark. The disadvantage of
equation (\ref{mss}) is that to simulate quarks with different
masses one should use two-component spinors at different position
in the same string. After we introduce gauge fields, link
variables which connect two different points, it seems that one
should insert nontrivial link variables between this two-component
spinors in the same string. We shall discuss this topic after
equation (\ref{d0}).

The kinetic part of the action  connects the $\Psi$'s associated
with two different strings trough the difference
$\varphi(x+a_\mu)-\varphi(x-a_\mu)$. For difference operators, we
define \beea \nabla_\mu \Psi(n)&=&{1\over
2} (\Psi(n+a_\mu)-\Psi(n-a_\mu)), \nonumber \\
\triangle_\mu\Psi(n)&=&{1\over
2}[\Psi(n+a_\mu)+\Psi(n-a_\mu)-2\Psi(n)]. \eea Then it is easy to
verify \beea
I^0&=&Tr[\bar{\Psi}(n)\gamma_0\bigtriangledown_0\Psi(n)]
-Tr[\lambda_0\lambda_5\bar{\Psi}(n)\gamma_5\bigtriangleup_0\Psi(n)]
\stackrel{\triangle}{=}I^0_1+I^0_2+I^0_3 \nonumber \\
&=&\sum\limits_x c_0(x)(-)^{x_0}\bar{\varphi}(n,x)
[\varphi(n,x+(-)^{x_0}\vec{x}_0)-
\varphi(n-(-)^{x_0}a_0,x+(-)^{x_0}\vec{x}_0)],
 \eea
where \beea I^0_1&=&{1\over 2}
Tr[\bar{\Psi}(n)\gamma_0\Psi(n+a_0)-
\lambda_0\lambda_5\bar{\Psi}(n)\lambda_5\Psi(n+a_0)] \nonumber
\\ &=&\bar{\varphi}(n,(1,0))\varphi(n+a_0,(0,0))+
\bar{\varphi}(n,(1,1))\varphi(n+a_0,(0,1)), \nonumber
\\ I^0_2&=&{1\over 2} Tr[-\bar{\Psi}(n)\gamma_0\Psi(n-a_0)-
\lambda_0\lambda_5\bar{\Psi}(n)\lambda_5\Psi(n-a_0)] \nonumber
\\ &=&-[\bar{\varphi}(n,(0,1))\varphi(n-a_0,(1,1))+
\bar{\varphi}(n,(0,0))\varphi(n-a_0,(1,0))],\nonumber \\
I^0_3&=&{1\over 2}\times 2 Tr[
\lambda_0\lambda_5\bar{\Psi}(n)\gamma_5\Psi(n)] \nonumber
\\ &=&\bar{\varphi}(n,(0,1))\varphi(n,(1,1))+
\bar{\varphi}(n,(0,0))\varphi(n,(1,0)) \nonumber \\ &&
-\bar{\varphi}(n,(1,1))\varphi(n,(0,1))-
\bar{\varphi}(n,(1,0))\varphi(n,(0,0)).
 \label{d0}\eea

Notice that interacting two-component spinors in $I^0_3$ belong to
the same string. That is, this interaction between spinors is self
interaction of the string spinor $\Psi$. Here we meet the same
puzzle as we treat mass problem in equation (\ref{mss}). For
interactions between two two-component spinors, if these two
two-component spinors belong to the same string, they correspond
to couplings between different components in the same string
spinor. We can regard the string spinor as two ordinary
four-component spinors at the same spatial-time point, then such
couplings are in fact the ones between components of ordinary
spinors in the same spatial-time point. We shall call such
couplings as self-couplings thereinafter. For instance, in
equation (\ref{mss}), the couplings are $\bar{\psi}(n)^i\psi_i(n)$
and $\bar{\psi}^i(n)\psi_j(n)(\lambda_0)^j_i$ while in equation
(\ref{d0}), the coupling is
$(\lambda_0\lambda_5)^i_j\bar{\psi}^j(n)\gamma_5\psi_i(n)$. It
should be noted here that after introduce gauge fields, such
coupling should be of independence with gauge configurations, as
required in continuum theory. In other words, it is not needed to
insert link variables in self-coupling. On the contrary, we should
insert gauge fields, or link variables in $I^0_1$ and $I^0_2$,
since they connect ordinary four-component spinors at different
points (strings).

This property of self-coupling makes that the value of $h$ in the
definition of $b_0$ and $b_1$ irrelevant. This can be seen that
all spinors connected by self-couplings, which are irrelevant to
covariant difference operators, will be interpreted as components
of the two  ordinary entangled four-component spinors, since they
belong to the same string. Therefore, we can choose $h\rightarrow
0$ in the simulation. After such choice, the four points in the
string, represented by $x$, will tend to the same point $n$. Since
that, we can always choose link variables, which connect these
four points in the same string, as unitary. In other words, the
insertion of gauge fields in self-coupling is not needed.

The needlessness of the insertion can also be seen as follows. In
the staggered action, quarks reflect the movement of string
spinors and they can not be regarded as point particles. In this
sense parameter $h$ determines the sizes of expansions of quarks.
On the contrary, in standard QCD, quarks are treated as point
particles and they have no inherent structure. Therefore, finite
parameter $h$ in staggered action presents the deviation from
standard QCD. At fixed lattice spacing, the more larger $h$ is,
the more sharper the deviation becomes.  However, at the fashion
scalar, this deviation should be very small, had this deviation
existed. Therefore, it seems that $h\ll 1$ in the simulation, due
to the experiments. This constraint makes that the gauge field
between points connected by self-coupling are weaker than the
gauge field between points connected by other interactions,
because of the asymptotic behavior of QCD.  In other words, the
insertion of gauge fields between point connected by self-coupling
is needless.

In this scheme, therefore, for each string, four $x$'s are only
used to distinguish different two-component spinors. In other
words, $x$'s belong to inner space and they are irrelevant to
spatial-time points.

Since two-component spinors belong to the same string spinor
(matrix) are in the same point, the cubic symmetry is restored.
For covariant derivative operators in spatial direction, the
situation is straightforward, then, \beea
I^i&=&Tr[\bar{\Psi}(n)\gamma_i\bigtriangledown_i\Psi(n)]
-Tr[\lambda_1\lambda_5
\bar{\Psi}(n)\lambda_5\otimes\tau_i\bigtriangleup_0\Psi(n)]
\stackrel{\triangle}{=}I^i_1+I^i_2+I^i_3 \nonumber \\
&=&\sum\limits_x c_1(x)(-)^{x_1}\bar{\varphi}(n,x)\tau_i
[\varphi(n,x+(-)^{x_1}\vec{x}_1)-
\varphi(n-(-)^{x_1}a_i,x+(-)^{x_1}\vec{x}_1)], \eea where
$\lambda_5\otimes\tau_i=i\gamma_i\gamma_0$ and \beea
I^i_1&=&{1\over 2} Tr[\bar{\Psi}(n)\gamma_i\Psi(n+a_i)-
\lambda_1\lambda_5\bar{\Psi}(n)i\gamma_i\gamma_0 \Psi(n+a_i)]
\nonumber
\\ &=&\bar{\varphi}(n,(0,1))\tau_i\varphi(n+a_i,(0,0))-
\bar{\varphi}(n,(1,1))\tau_i\varphi(n+a_i,(1,0)), \nonumber
\\ I^i_2&=&{1\over 2} Tr[-\bar{\Psi}(n)\gamma_i\Psi(n-a_i)-
\lambda_1\gamma_5\bar{\Psi}(n)i\gamma_i\gamma_0\Psi(n-a_i)]
\nonumber
\\ &=&\bar{\varphi}(n,(1,0))\tau_i\varphi(n-a_i,(1,1))-
\bar{\varphi}(n,(0,0))\tau_i\varphi(n-a_i,(0,1)),\nonumber \\
I^i_3&=&{1\over 2}\times 2 Tr[
\lambda_1\lambda_5\bar{\Psi}(n)i\gamma_i\gamma_0\Psi(n)] \nonumber
\\ &=&\bar{\varphi}(n,(1,1))\tau_i\varphi(n,(1,0))-
\bar{\varphi}(n,(1,0))\tau_i\varphi(n,(1,1)) \nonumber \\ &&
-\bar{\varphi}(n,(0,1))\tau_i\varphi(n,(0,0))+
\bar{\varphi}(n,(0,0))\tau_i\varphi(n,(0,1)). \label{di} \eea

If we define $\varphi(n,(\pm2,x_1))=\varphi(n\pm a_0,(0,x_1))$,
$I^0$ can be written in a more symmetry form, \bee I^0=
\sum\limits_{x\in
n}\bar{\varphi}(n,x)c_0(x)(\varphi(n,x+\vec{x}_0)-
\varphi(n,x-\vec{x}_0)). \label{endaction1} \ee Similarly, if we
define $\varphi(n,(x_0,\pm 2))=\varphi(n\pm a_i,(x_0,0))$ in
$I^i$, $I^i$ can also be written as \bee I^i= \sum\limits_{x\in
n}\bar{\varphi}(n,x)c_1(x)
\tau_i(\varphi(n,x+\vec{x}_1)-\varphi(n,x-\vec{x}_1)).
\label{endaction} \ee

This two equations are very similar to the standard staggered
action. Utilizing equation (\ref{mss})\,-\,(\ref{endaction}), one
can easily construct action, in which there are two fermions with
different masses. In the form of string spinors, the action is as
\bee \label{action} S=\sum\limits_{n} Tr[m_0\bar{\Psi}(n)\Psi(n)+
m_1\bar{\Psi}(n)\Psi(n)\lambda_0+\bar{\Psi}(n)\gamma_\mu
\bigtriangledown_\mu\Psi(n)- \lambda_{min(1,\mu)}\lambda_5\Psi(n)
\lambda_5\otimes\bar{\tau}_\mu\triangle_\mu \Psi(n)], \ee where
$\bar{\tau}=({\bf 1},\vec{\tau})$.

In this scheme one should choose $h\rightarrow 0$ at present
scalar. However, it is also interesting to set $h\simeq 1$ and put
insertion of gauge fields between different points connected by
couplings, including self-couplings, even only for maturity of the
theory consideration or for the future lattice simulations. At
this case, especially for $h=1$, we define link variables on all
the half-integer points and integer points. All the coupling,
including self-coupling, can be categorized into two types. The
first is the coupling between points with different time but with
the same spatial, such as coupling between $(n,(0,0))$ and
$(n,(1,0))$. We need insert only one link variable between these
point.  The second is the coupling between points with different
spatial but with the same time, for instance, coupling between
points $(n,(0,0))$ and $(n,(0,1))$. At this case we need insert
three link variables between these points. However, there are
eight paths between this points, in other words, we have eight
different insertions of gauge fields between these points. One
should make an average between these insertions. This is just a
smearing process, which is commonly used in many lattice
simulations. Surely one should also perform a tadpole improvement
here, since the insertion of three link variables leads to large
tadpole correction.

\section{Basic dynamics of string spinors}

When $h\rightarrow 0$, we have a string structure in each "point"
(string) in this scheme. In each "point" of QCD, there is a
curling string, which connected four two-component spinors. Since
the string is very small, it seems that there exists interaction
between these spinors. However, the force should decay very
sharply with the increase of distance, for the interaction between
spinors on string is adjoining. An extreme case is that we choose
$h=0$ directly. At this time the string is living on an extra
dimension.

This interaction is not QCD. We first notice there is a basic
symmetry of this interaction. That is, for $I^0$ and $I^i$, there
is a symmetry under the following discrete transformation, \beea
\Psi(n)&\rightarrow&i\gamma_5\Psi(n)\lambda_5, \nonumber \\
\bar{\Psi}(n)&\rightarrow&i\lambda_5\bar{\Psi}(n)\gamma_5. \eea
Notice that the mass splitting term in (\ref{mss}) is also
invariant under this transformation. We are able to rewrite this
transformation in the next form, \beea \varphi((1,0))&\rightarrow&
-\varphi((1,0)),\,
\varphi((0,1))\rightarrow -\varphi((0,1)),\, \nonumber \\
\bar{\varphi}((0,0))&\rightarrow& -\bar{\varphi}((0,0)),\,
\bar{\varphi}((1,1))\rightarrow -\bar{\varphi}((1,1)),
\label{con1} \eea  with other variables, $\varphi((1,1))$,
$\varphi((0,0))$, $\bar{\varphi}((1,0))$, and
$\bar{\varphi}((0,1))$, invariance. Or \beea
\varphi((1,1))&\rightarrow& -\varphi((1,1)),\,
\varphi((0,0))\rightarrow -\varphi((0,0)),\, \nonumber \\
\bar{\varphi}((0,1))&\rightarrow& -\bar{\varphi}((0,1)),\,
\bar{\varphi}((1,0))\rightarrow -\bar{\varphi}((1,0)),
\label{con2} \eea with other variables invariance.

At both case ($h\rightarrow 0$ and $h=0$), the string spinor,
$\Psi$, has a inherent structure. To describe this structure one
should find the dynamical variables of the string and spinors and
the Lagrangian of dynamical variables. One may choose the four
tastes described as $\varphi(x)$ for fermion freedom. To study
dynamics of fermion one should insert interaction between them,
which is a gauge interaction determined by boson freedom. Notice
the role of $c_\mu$ in equations (\ref{endaction1}) and
(\ref{endaction}), one may consider that $c_\mu$ just reflects
this interaction. In other words, this interaction is possibly
described by $U(1)$ theory on the string, the topology of which is
also $U(1)$.

 The Lagrangian, which describes movement dynamics of $\varphi(x)$'s
and $A_\rho$'s ($A_\rho$ are defined through link variables,
$c_\rho=e^{i\int_{path} A_\rho}$), should satisfy the constraint
shown in equations (\ref{con1}) and (\ref{con2}). This means there
exist only adjoining interactions between this four tastes/spinors
on the string. The interaction which connects tastes on diagonal
points, such as interaction $\bar{\varphi}((0,0))\varphi((1,1))$,
can not occur.

Since $c_\rho=e^{i\int_{path} A_\rho}$ is a link variable which
connects the adjoining tastes, it seems that the value of $c_\rho$
is arbitrary, provided it satisfies $c_\rho(x) c^*_\rho(x)=1$. It
is obvious that one can always choose a gauge to make $c_\rho$
satisfy equation (\ref{c01}). Therefore, the choice of $c_\rho$ in
equation (\ref{c01}) can be regarded as a special gauge fixing on
$c_\rho$. However, a gauge independent variable \bee c_{string}
=c_1((0,0))c_0((0,1))c_1^*((1,0))c^*_0((0,0))=-1 \ee for string
implies that boson freedom $A_\rho$ is not arbitrary fluctuating.
In fact, this identity implies that for each string (plaquette) we
have \bee \oint d{\bf l}\cdot{\bf A}\equiv (2
k+1)\pi,\label{wind}\ee where $k$ is a integer. Since $A_\rho$
plays as a phase factor of $c_\rho$ and the couplings between
different $\varphi$ is not $A_\rho$ but $c_\rho$, different $k$
corresponds to the same dynamics of string. We choose $k=0$ here.
Therefore, there exists a constraint on boson freedom $A_\rho$.

Suppose the string is living on an extra dimension. Since the
interaction is $U(1)$ gauge theory defined on manifold $U(1)$, it
is interesting to study the topology behavior of $A_\rho$. Then it
is natural to define winding number, which is topological
invariant under infinitesimal gauge transformation, in the
manifold $U(1)$ as \bee n={1\over 2\pi}\oint d{\bf l}\cdot {\bf
A}={1\over 4\pi}\int\limits_{string} ds\,
\epsilon_{\rho\lambda}F_{\rho\lambda}, \ee where the second
integrand is over the area encircled by the string in inner space
and $F_{\rho\lambda}=\partial_\rho A_\lambda-\partial_\lambda
A_\rho$, $\rho,\,\lambda=1,\,2$. The choice of $k=0$ in equation
(\ref{wind}) means that $n={1\over 2}$. There is no connection
between $c_\rho$ and topological quantity, winding number, in
standard staggered action, whereas in our scheme there exists a
deep connection between them. We think that the dynamics of the
string, especially the nontrivial topology behavior, reveals that
it is worth studying furthermore. For instance, whether there
exists a relation between nontrivial topology of the string and
the broken of chiral symmetry and how the relation occurs, if the
existent is positive.

There is subtlety in rotation. In standard staggered action they
rotation symmetry is very complex. On lattice this symmetry
includes not only the usual hypercubic group, the subgroup of the
continuous rotation group, but also the spin-isospin
mixing\ct{zhang}. The hypercubic symmetry guarantees that the
rotation symmetry is restored  at the limit lattice spacing
$a\rightarrow 0$ (Suppose we only give our attention to the
rotation in spatial), while the other part of the symmetry, the
spin-isospin symmetry, does not occur in standard QCD. There is a
small symmetry in our action, (\ref{action}). Particularly, it
seems that the system is not invariant under transformations in
cubic group. However, as revealed in reference \cite{zhang},
$c_\rho$ is spin-zero field. We furthermore consider that the
string, represented by series of $c_\rho$'s, is also invariant
under the rotation, including its subgroup, cubic group. It is
more obvious if we think the string is living on extra dimension.
Since the string, especially, the orient of string, is invariant
under rotation, the relative positions of the four tastes are also
invariant, that is, there is no spin-isospin mixing under cubic
transformation in our action. Therefore, the only  symmetry of our
action is the cubic group, which excludes the spin-isospin mixing.
This can also be seen in action (\ref{action}). If the cubic
rotations exclude the interchange of spin-isospin, that is,
$\psi^1$ can not change to $\psi^2$, or vis versa, under cubic
transformation( This means that, $\psi^1$ and $\psi^2$ transform
as a independent four-component spinor under cubic
transformation), the system is invariant under the cubic group.
The cubic group guarantees that the continuous rotation invariant
is restored in the continuum limit. In this scheme we have
discarded the spin-isospin mixing symmetry which does not occur in
standard QCD.

In summary, each point of QCD is described as a string, on which
live four fermion tastes. Boson freedom, $A_\rho$ (or $c_\rho$),
and fermion freedom, four fermion tastes $\varphi(x)$, make up of
the complete variables of string. But fermion freedom and boson
freedom themselves are not observable quantity. Observable
quantities are described as the dynamics of these freedoms. For
instance, the two types of quarks are determined by the dynamics
of the variables.

\section{Summary}
In this note we show a quasi-staggered action, which preserves the
cube symmetry. This action regards that the point of QCD has a
inherent structure. Each point of QCD is in fact a string, which
connects four interacting tastes with spin $1\over 2$. However,
these tastes themselves are not observed spinors. The observed
fermions (quarks) are the eigen-models of the dynamics of the
tastes. One byproduct is that quarks should be in pair in this
action. However, it is not needed to require quarks, which occur
in pair, be degenerate, that is, masses of the pair of quarks can
be different. This is a significant property of this action. The
other byproduct is that the spin-isospin interchange symmetry does
not occur in this scheme.

The string structure of the "point" of vacuum in this action is an
amusing picture. We possibly meet a bridge between staggered
action in QCD and a more modern physics, string or superstring
theory\ct{wt}. It is interesting to compare the similarities and
differences between the picture of string adopted here and that in
standard string theory.



\begin{thebibliography}{99}

\bibitem{nn}
R. Gupta, hep-lat/9808027; H.B. Nielsen and M. Ninomiya, Nucl.
Phys. {\bf B185} (1981)20; Erratum Nucl. Phys. {\bf B195}
(1982)541; Nucl. Phys. {\bf B193} (1981)173.


\bibitem{wilson}
K.G.Wilson, in New Phenomena in Subnuclear Physics, ed.A.Zichichi
(Plenum Press, 1977).



\bibitem{overlap}
Y. Kikukawa, R. Narayanan and H. Neuberger, Phys. Rev.
{\bf D57}(1998) 1233, hep-lat/9705006; R.
Narayanan hep-lat/9707035.

\bibitem{slac}
S.D. Drell, M. Weinstein and S. Yankielowicz, Phys. Rev. {\bf D14}(1976), 487;
T. Sugihara, hep-lat/0310061.

\bibitem{kgt}
J. Kogut and L. Susskind, Phys. Rev. {\bf D11} (1975) 395; L.
Susskind, Phys. Rev. {\bf D16} (1977) 3031.

\bibitem{stagger}
J.Kogut and L.Susskind, Phys. Rev. {\bf D11}
 (1975)395; T. Banks, J.Kogut and L.Susskind, Phys. Rev. {\bf D13}
(1996)1043; L. Susskind, Phys. Rev. {\bf D16} (1977) 3031.

\bibitem{np1}
F. Gliozzi, Nucl.Phys. {\bf B204} (1982) 419.

\bibitem{rooting}
S. D\"{u}rr, C. Hoelbling and U. Wenger, Phys. Rev. {\bf D70}
(2004) 094502; S. D\"{u}rr, hep-lat/0509026.


\bibitem{wt}
M.B. Green, J.H. Schwarz and E. Witten, Superstring Theory,
Combridge university Press, USA, (1986) 18.

\bibitem{zhang}
G. Parisi, Zhang Yi-cheng, Nucl.Phys. {\bf B230} (1984) 97.


\end{thebibliography}
\end{document}